\documentclass{ieeeaccess}
\usepackage{amsmath,amssymb,amsfonts}
\usepackage{algorithmic}
\usepackage{graphicx}
\usepackage{textcomp}
\usepackage{amsmath}
\usepackage{hyperref}
\usepackage{caption}
\usepackage[url= false,
sorting=none,
citestyle=numeric-comp
]{biblatex}
\usepackage{subcaption}
\usepackage{braket}
\usepackage{float}
\usepackage{multirow}

\DeclareMathOperator*{\argmax}{arg\,max}
\DeclareMathOperator*{\argmin}{arg\,min}

\def\BibTeX{{\rm B\kern-.05em{\sc i\kern-.025em b}\kern-.08em
    T\kern-.1667em\lower.7ex\hbox{E}\kern-.125emX}}
\begin{document}
\doi{}

\title{Empirical Study of Observable Sets in Multiclass Quantum Classification}

\author{\uppercase{Paul San Sebastian Sein}\authorrefmark{1,2}
\uppercase{Mikel Cañizo \authorrefmark{2}, Román Orús\authorrefmark{3,4,5}}.
}
\address[1]{Ikerlan Technology Research Centre, Basque Research and Technology Alliance (BRTA),
Paseo J.M. Arizmediarrieta 2, E-20500, Arrasate-Mondragon, Spain}
\address[2]{University of the Basque Country/Euskal Herriko Unibertsitatea-UPV/EHU}
\address[3]{Donostia International Physics Center, Paseo Manuel de Lardizabal 4, E-20018 San Sebastian, Spain}
\address[4]{Multiverse Computing, Paseo de Miramon 170, E-20014 San Sebastian, Spain}
\address[5]{Ikerbasque Foundation for Science, Maria Diaz de Haro 3, E-48013 Bilbao, Spain}


\markboth
{P. San Sebastian \headeretal: Empirical Study of Observable Sets in Multiclass Quantum Classification }
{P. San Sebastian  \headeretal: Empirical Study of Observable Sets in Multiclass Quantum Classification}

\corresp{Corresponding author: Paul San Sebastian Sein (email: psansebastian@ikerlan.es).}

\begin{abstract}
Variational quantum algorithms have gained attention as early applications of quantum computers for learning tasks. In the context of supervised learning, most of the works that tackle classification problems with parameterized quantum circuits constrain their scope to the setting of binary classification or perform multiclass classification via ensembles of binary classifiers (strategies such as \textit{one versus rest}). Those few works that propose native multiclass models, however, do not justify the choice of observables that perform the classification. This work studies two main classification criteria in multiclass quantum machine learning: maximizing the expected value of an observable representing a class or maximizing the fidelity of the encoded quantum state with a reference state representing a class. To compare both approaches, sets of Pauli strings and sets of projectors into the computational basis are chosen as observables in the quantum machine learning models. Observing the empirical behavior of each model type, the effect of different observable set choices on the performance of quantum machine learning models is analyzed in the context of Barren Plateaus and Neural Collapse. The results provide insights that may guide the design of future multiclass quantum machine learning models.
\end{abstract}


\titlepgskip=-15pt

\maketitle


\section{Introduction}
\begin{figure*}[h] 

\begin{subfigure}{0.5\textwidth}
\includegraphics[width=0.9\linewidth, height=6cm]{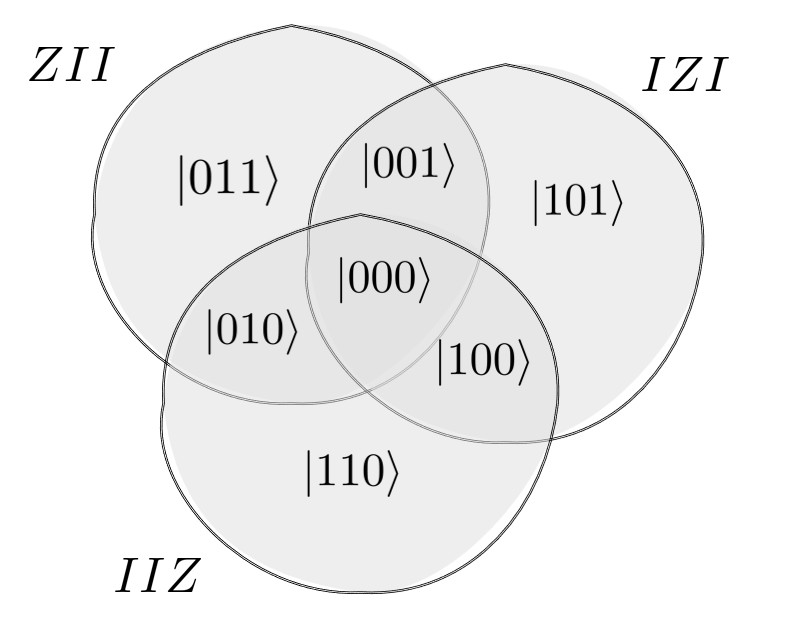} 
\caption{Overlap between positive eigenspaces of Pauli strings.}
\label{fig:eigenspaces}
\end{subfigure}
\begin{subfigure}{0.5\textwidth}
\includegraphics[width=0.9\linewidth, height=6cm]{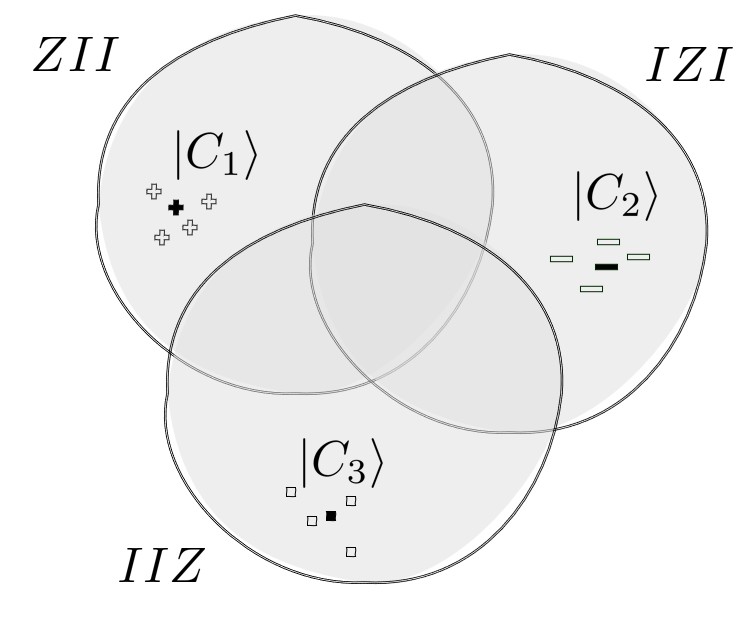}
\caption{Arrangement of class mean state vectors $|C_i\rangle$ forming an ETF.}
\label{fig:eigencentroids}
\end{subfigure}

\caption{Schematic representation of positive eigenspaces of Pauli strings.}
\label{fig:egigenss}
\end{figure*}

\label{sec:introduction}
\PARstart{Q}{uantum} machine learning (QML) \cite{cerezo_variational_2021, biamonte_quantum_2017, schuld_machine_2021, mitarai_quantum_2019, kopczyk_quantum_nodate}, has emerged as a promising frontier at the intersection of quantum computing and data-driven artificial intelligence, seeking to exploit quantum phenomena for enhanced learning performance. Within this framework, supervised learning with parameterized quantum circuits (PQCs) has attracted particular attention due to its potential to represent complex decision boundaries using a compact number of quantum resources \cite{schuld_supervised_2018}. PQC-based models encode classical data into quantum states and optimize tunable gate parameters through classical feedback loops, forming hybrid quantum–classical architectures that mirror the structure of neural networks while leveraging quantum phenomena. These models have demonstrated the capacity to perform classification tasks on low-dimensional datasets, offering insights into how quantum representations may scale for more complex problems \cite{huang_power_2020}. Despite hardware limitations and challenges such as \textit{Barren Plateaus} \cite{larocca_barren_2025}, circuit expressibility \cite{schuld_effect_2020}, and generalization performance \cite{caro_generalization_2022, gil-fuster_understanding_2024}, ongoing research continues to refine training strategies, encoding schemes, and cost functions tailored to quantum hardware. The study of PQC-based supervised learning not only deepens our understanding of quantum representation power but also contributes to developing practical quantum classifiers capable of outperforming or complementing their classical counterparts in the noisy intermediate-scale quantum (NISQ) era.

 Most of seminal works in quantum machine learning, however, constrain themselves to binary classification problems or approach multiclass classification through ensembles of binary classifiers. These ensembles can follow strategies such as \textit{one versus rest} (OvR) or \textit{ one versus one} (OvO) \cite{kang_constructing_2015,hong_probabilistic_2008,silver_quilt_2022,bishwas_all-pair_2018}, which are widely employed in the classical machine learning context. Nevertheless, these methods require a number of classifiers that scale linearly in the case of OvR and quadratically for OvO. Training a high number of classification heads can be expensive in the current quantum computing era, considering the quantum noise, the Barren Plateaus and technical implementation problems in the state-of-the-art quantum devices. For this reason, it is convenient to design a quantum machine learning model that is natively able to discriminate between multiple classes of data. 

There are several preliminary works that propose variational quantum circuits for multiclass classification \cite{stein_quclassi_2022,ding_scalable_2024,chalumuri_hybrid_2021}, but just few of them pay special attention to the choice of the set of observables and the impact of this choice in the performance of the model. For instance, \cite{yun_projection_2022} distinguishes between \textit{Positive Operator-Valued Measure} (\textit{POVM}) and \textit{Projection-valued Measure} (\textit{PVM}) observables and combine both in different steps of a general classification algorithm. Authors in \cite{lee_variational_2024} propose a parameterized quantum circuit measured by \textit{POVM}s that perform supervised classification following a state discrimination approach. \cite{nghiem_unified_2021} proposes two different approaches to perform multiclass classification: the \textit{implicit} and the \textit{explicit} approaches. The first approach, although conceptually simple and powerful, requires a QRAM for its efficient implementation, whereas the second approach generalizes the method followed by \textit{quantum metric learning} \cite{lloyd_quantum_2020}. Finally, and closely related to our work, \cite{du_problem-dependent_2023} proves that multiclass QML models with a set of orthogonal measurements can reach Neural Collapse and propose an interesting framework to study the generalization of QML models.

Most of the cited works tackle multiclass classification using one of two main strategies. The first assigns a specific observable to each class and then labels a datapoint according to the class whose observable has the largest (or smallest) expected value for the datapoint’s evolved quantum state. The second defines a reference quantum state for each class and then assigns a datapoint to the class whose reference state is closest to the datapoint’s evolved quantum state. Both methods are closely related, but they hide subtle differences that can influence the performance of the model (see Sec. \ref{sec:model_description}.

In this work, we study and empirically compare these two approaches. For the first one, we consider a set of Pauli-string observables, and for the second, we use a set of computational basis states as reference states. First, we investigate how the locality of each set of observables affects the concentration of the loss, a phenomenon related to Barren Plateaus. We then train classification models on several datasets and analyze how the choice of observable influences classification performance. In addition to standard classification metrics, we also study the phenomenon of Neural collapse, a well-known effect that arises in deep learning models during their terminal phase of  \cite{papyan_prevalence_2020, lu_neural_2022}.

In summary, the main contributions of this work are as follows:
\begin{itemize}
\item A review and characterization of Barren Plateaus and Neural Collapse in relation to the choice of observables.
\item An experimental analysis of how different observable choices affect model performance, loss-function concentration, and convergence toward Neural Collapse.
\item The derivation of empirically grounded conclusions, consistent with theoretical expectations, that can inform the design of future multiclass quantum machine learning models.
\end{itemize}

This paper is organized as follows: section \ref{sec:preliminaries} introduces the background of the concepts that are necessary to understand the numerical results; starting from the classification with parameterized quantum circuits (sec. \ref{sec:classification_with}), to basic notions of Pauli strings (sec. \ref{sec:paulistrings}), projective measurements (sec. \ref{sec:projective_measurements}), Barren Plateaus (sec. \ref{sec:BPs}) and Neural Collapse (sec. \ref{sec:NC}). Afterwards, the different types of QML models trained during work are defined in sec. \ref{sec:model_description} and details about datasets and the experimental setting are given in sec.\ref{sec:setting}. Numerical results an their interpretation are presented in sec. \ref{sec:numerical_results}, and finally, the manuscript finishes summarizing the conclusions and future steps in sec. \ref{sec:conclusion}.

\section{Preliminaries} \label{sec:preliminaries}
\subsection{Classification with parameterized quantum circuits} \label{sec:classification_with}
 
 We first define the classification problem and the notation used during the work. In the context of supervised learning, a set of $N$ data points $\mathbf{x}_i \in \mathbb{R}^m$ is provided along with their respective labels $y_i\in \{1,2,\dots, K\}$. Unlike in \textit{multilabel} classification, each data point belongs to a single categorical class. 

 Data points are encoded in a PQC by a unitary $U(\mathbf{x_i}, \mathbf{\theta})$ and are represented as quantum states living in a $2^{n_q}$ - dimensional Hilbert space, where $n_q$ is the number of the qubits in the quantum circuit. By tuning the variational parameters $\mathbf{\theta}$ of the quantum circuit, one can modify the quantum state $|\psi(\mathbf{x_i}, \mathbf{\theta})\rangle$ that represents a data point (named \textit{feature state} during this work)\footnote{Only pure quantum states are considered in this work.}. A quantum machine learning model is typically defined as the expected value of an observable ${O}$ for a given evolved quantum state \cite{du_quantum_2025}:
 \begin{equation}
 f(\mathbf{x}_i,\theta) = \langle {O}\rangle_{\psi(\mathbf{x_i}, \mathbf{\theta})} = \langle\psi_0|U^{\dagger}(\mathbf{x}_i, \mathbf{\theta}){O}U(\mathbf{x}_i, \mathbf{\theta})|\psi_0\rangle,
 \end{equation}

 being $|\psi_0\rangle$ a initial quantum state before encoding the data point. It has been proved that, when the data is encoded redundantly following the so-called \textit{data re-uploadong} method \cite{perez-salinas_data_2019}, the quantum model $f$ can be expressed as a partial Fourier series and can be a Universal Function Approximator \cite{schuld_effect_2020}. This analytical framework has been helpful to obtain theoretical results about the power of QML models or their simulability \cite{landman_classically_2022}.

For multiclass classification, we define a vector $\mathbf{f}_{\mathbf{\theta}}$ with expected values of observables ${O}_i$ associated to each class:
\begin{equation}
\label{eq:expval_vector}
\mathbf{f}(\mathbf{x_i},\theta) = \big[\langle {O}_1\rangle_{\psi(\mathbf{x_i}, \mathbf{\theta})},\langle {O}_2\rangle_{\psi(\mathbf{x_i}, \mathbf{\theta})},\dots,\langle {O}_K\rangle_{\psi(\mathbf{x_i}, \mathbf{\theta})} \big].
\end{equation}

 We then choose a function of $\mathbf{f}$ as a classification criterion. For example, a data point can be labeled as 

 \begin{equation}
     \hat{y}(\mathbf{x_i}, \theta) = \argmax_{k=1,\dots,K} \mathbf{f}_k(\mathbf{x_i},\mathbf{\theta}).
 \end{equation}

The training of the quantum model is performed by a classical optimizer, in order to minimize a loss function $\ell$ that quantifies the dissimilarity between the predicted and true labels:

\begin{equation}
    \theta^* = \argmin_{\theta} \frac{1}{N} \sum_i^N \ell(y_i, \hat{y}(\mathbf{x_i}, \theta)).
\end{equation}

\subsection{Pauli Strings}\label{sec:paulistrings}
Pauli strings are defined as tensor products of local Pauli matrices \cite{dion_efficiently_2024}:

\begin{align}
    {P} = \bigotimes_{i=1}^{n_q} {\sigma}_{i} && \textrm{where}  && {\sigma}_i \in \{X,Y,Z,I\}.
\end{align}

These are well studied operators, since they form an orthonormal basis for the operator space on $n_q$ qubits. When measuring a quantum computer, the evolved state vector is projected into the basis of eigenvectors of a certain operator, by convention into the eigen-basis of Pauli-Z operators acting in each qubit \cite{crawford_efficient_2021}, also known as the computational basis. Therefore, the quantum circuit will sample a computational basis state with a probability following the Born's rule \cite{pais_max_1982}, and the expected value of an observable with respect to a certain quantum state is given by:
\begin{equation}
\label{expval}
    \langle {O}\rangle_{\psi} = \sum_i p(\lambda_i)\lambda_i
= \sum_i |\langle \psi | \lambda_i \rangle |^2 \lambda_i,
\end{equation}

where $\lambda_i$ are the eigenvalues of $O$, and $p(\lambda_i)$ and $|\lambda_i\rangle$ are their associated eigenvectors and probabilities, respectively \cite{nielsen_quantum_2010, rieffel_quantum_2011}. In the case of Pauli strings, $\lambda_i \in \{-1,1\}$, and in consequence, $\langle P\rangle_\psi \in [-1,1]$, for all possible quantum states. 

When a pair of Pauli strings doesn't commute, the expected value of both operators can't be computed by sampling in the same basis as in (\ref{expval}), since there is no basis that diagonalize both operators simultaneously. For this reason, in certain applications, it is convenient to choose mutually-commuting Pauli string sets, in order to reduce the sampling complexity \cite{crawford_efficient_2021}. There exist $4^{n_q}$ distinct $n_q$-long Pauli strings, but the maximum set of mutually-commuting Pauli strings contains $2^{n_q}$ elements. 

The locality of an operator is the number of qubits on which it acts non-trivially, i.e. the number of local non-identity operators in the tensor product. As it will be explained in \ref{sec:BPs}, the locality of an observable can affect in the trainability of a QML model, but it is also an important property in quantum tomography and shadowing algorithms \cite{huang_predicting_2020}.

\subsection{Projective Measurements}\label{sec:projective_measurements}
A POVM can be defined as a set of operators $\{E_m\}$ such that each operator $E_m$ is positive and they fulfill the completeness relation $\sum_m E_m = I$ \cite{nielsen_quantum_2010}. During this work, a special POVM is considered, where each operator is a projector of the form $\Pi_m = \ket{m}\bra{m}$ and they all follow $\Pi_m\Pi_{m'}=\Pi_m\delta_{mm'}$.

Working with pure states, computing the expected value of a projector
\begin{equation}
    \langle\Pi_m\rangle_\psi= \braket{\psi|m}\braket{m|\psi} = |\braket{\psi|m}|^2
\end{equation}
 is equivalent to compute the fidelity between $\ket{\psi}$ and $\ket{m}$. Experimentally, this quantity can be measured either by performing a change of basis from the computational basis into a basis containing $\ket{m}$ and sampling in the new basis, or by performing the \textit{swap test} \cite{schuld_supervised_2018} using ancillary registers.
 
\subsection{Barren Plateaus} \label{sec:BPs}
Barren Plateaus have been identified as one of the main limitations of variational quantum algorithms and have been extensively studied in the last years \cite{holmes_barren_2021, mcclean_barren_2018, larocca_barren_2025, ortiz_marrero_entanglement-induced_2021}. The term Barren Plateau refers to the exponential concentration of the loss landscape around its mean value as a function of the number of qubits in the parameterized quantum circuit. As a consequence, training in such a landscape becomes exponentially harder, since the magnitudes of the gradients also decay exponentially. Formally, a Barren Plateau is described by 

\begin{equation}
    \operatorname{Var}_{\theta}[l_\theta(\rho,O)] \; \textrm{or} \,\operatorname{Var}_{\theta}[\partial_\mu l_\theta(\rho,O)] \in \mathcal{O}\big(\frac{1}{b^{n_q}}\big),
\end{equation}

where $\partial_\mu l_\theta(\rho,O) = \frac{\partial l_\theta(\rho,O)}{\partial \theta_\mu}$, $\forall \theta_\mu\in\theta$, being $\rho$ a general quantum state, and $b >1$ \cite{larocca_barren_2025}. Different sources of Barren Plateaus have been identified, such us the expressibility of the parameterized quantum circuit \cite{holmes_connecting_2022, ortiz_marrero_entanglement-induced_2021}, noise \cite{wang_noise-induced_2021} or the locality of the cost function \cite{cerezo_cost_2021, uvarov_barren_2021}, but recent results gather all sources in a compact, exact and elegant Lie-algebraic expression \cite{ragone_lie_2024}:

\begin{equation}
\label{BPmaster}
 \operatorname{Var}_{\theta}[l_\theta(\rho,O)] = \frac{\mathcal{P}_{\mathfrak{g}}(\rho)\mathcal{P}_{\mathfrak{g}}(O)}{\textrm{dim}(\mathfrak{g})},
\end{equation}

where $\mathfrak{g}$ is the Dynamical Lie Algebra (DLA) of the parameterized quantum circuit and $\mathcal{P}_{\mathfrak{g}}(H)$ is the $\mathfrak{g}$-purity of a Hermitian operator

\begin{equation}
 \mathcal{P}_{\mathfrak{g}}(H) = \operatorname{Tr}\big[H_{\mathfrak{g}}^2\big] = \sum_{j=1}^{\textrm{dim}(\mathfrak{g})}|\operatorname{Tr}\big[B_j^\dagger H \big]|^2, 
\end{equation}
being $B_j$ vectors of the orthogonal basis of $\mathfrak{g}$. From the result in \ref{BPmaster}, we can identify the different sources that contribute to the concentration of the loss, being the dimension of the DLA the main source, since in typical ansätze used in many QML applications, this quantity grows exponentially $\textrm{dim}(\mathfrak{g})\in \mathcal{O}(4^{n_q})$. However, even if the DLA of the chosen ansatz lives in a polynomialy big space, the variance of the loss can still decay exponentially through the purity terms \cite{larocca_barren_2025}. 


\begin{figure*}[h!] 
\centering

\begin{minipage}[t]{0.3\textwidth}
    \vspace{0pt}
    \centering
    \includegraphics[width=\linewidth]{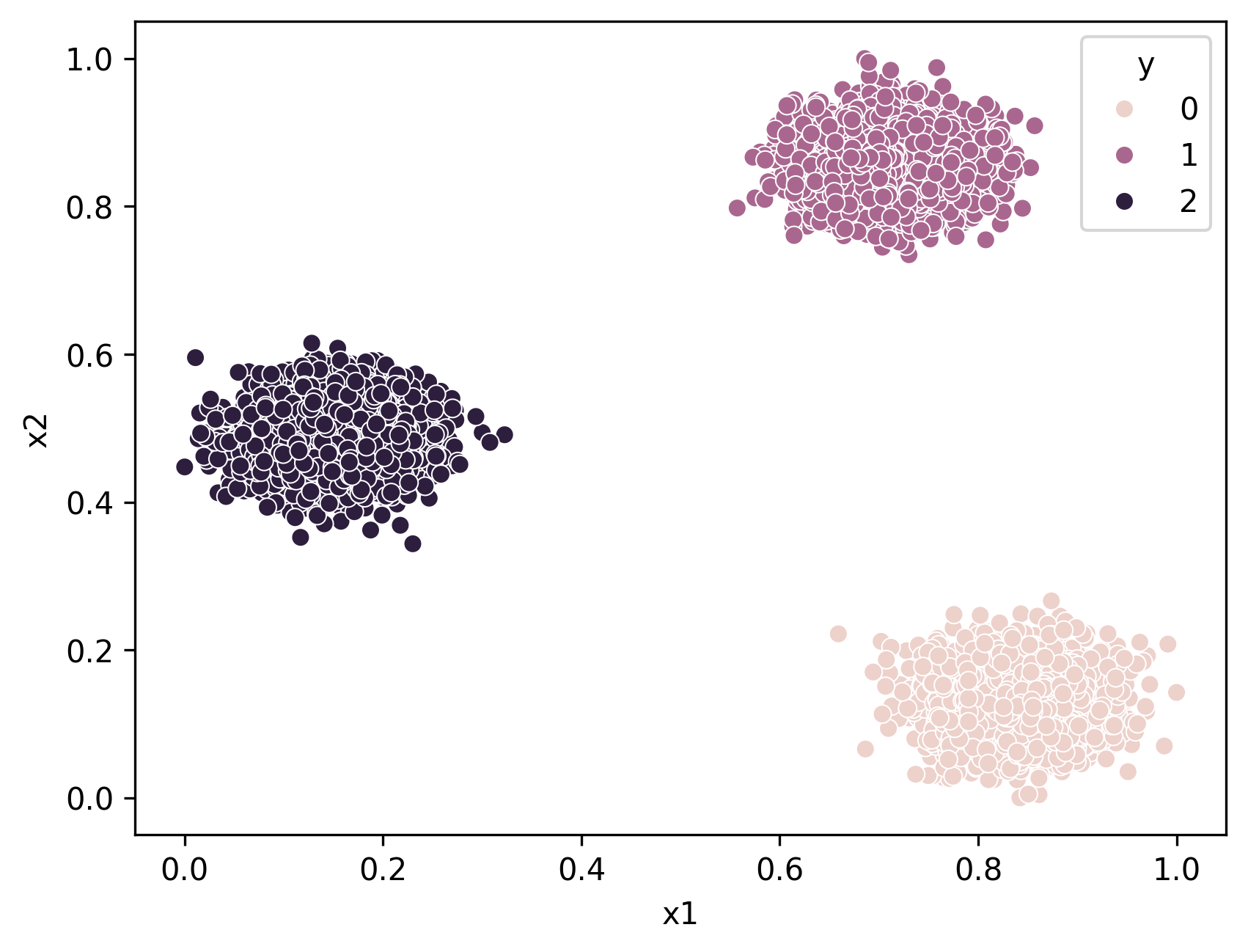}
    \caption{Visual representation of the dataset \textit{Blobs3}.}
    \label{fig:blobs3}
\end{minipage}%
\hfill
\begin{minipage}[t]{0.3\textwidth}
    \vspace{0pt}
    \centering
    \includegraphics[width=\linewidth]{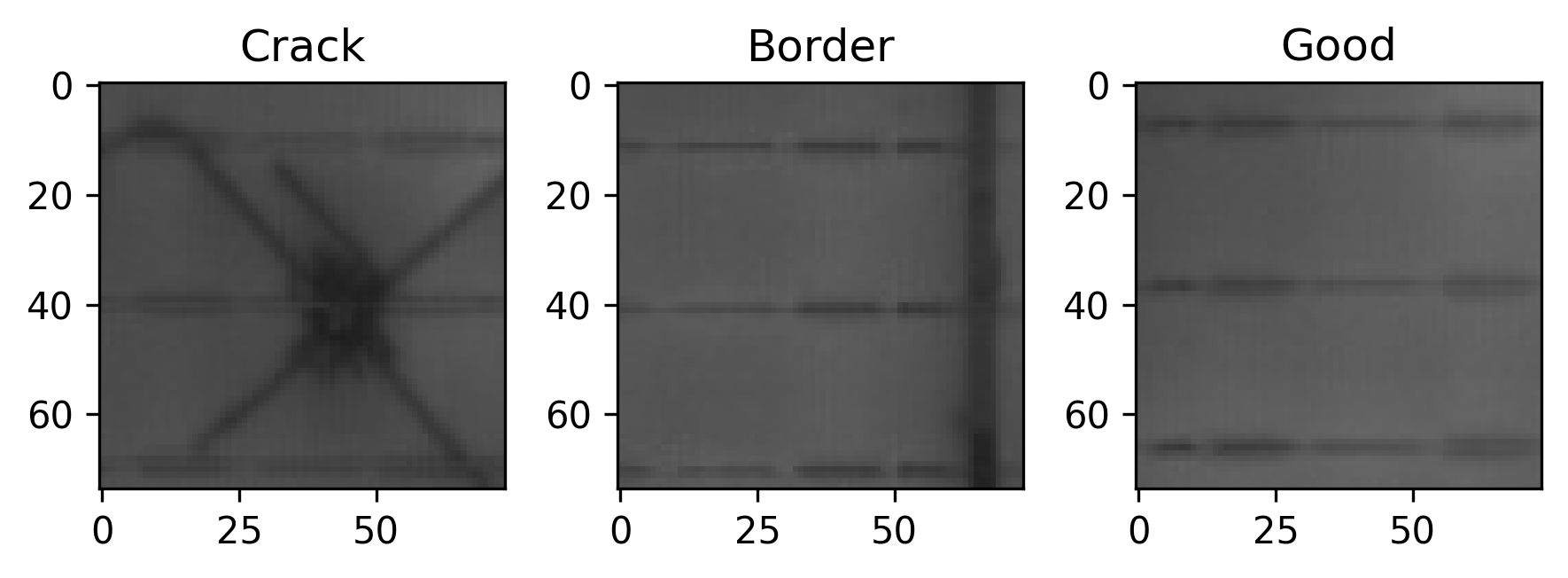}
    \caption{Visualization of instances of each class of the dataset \textit{Panels}.}
    \label{fig:panels}

    \par\vspace{0.5em}
    \includegraphics[width=\linewidth]{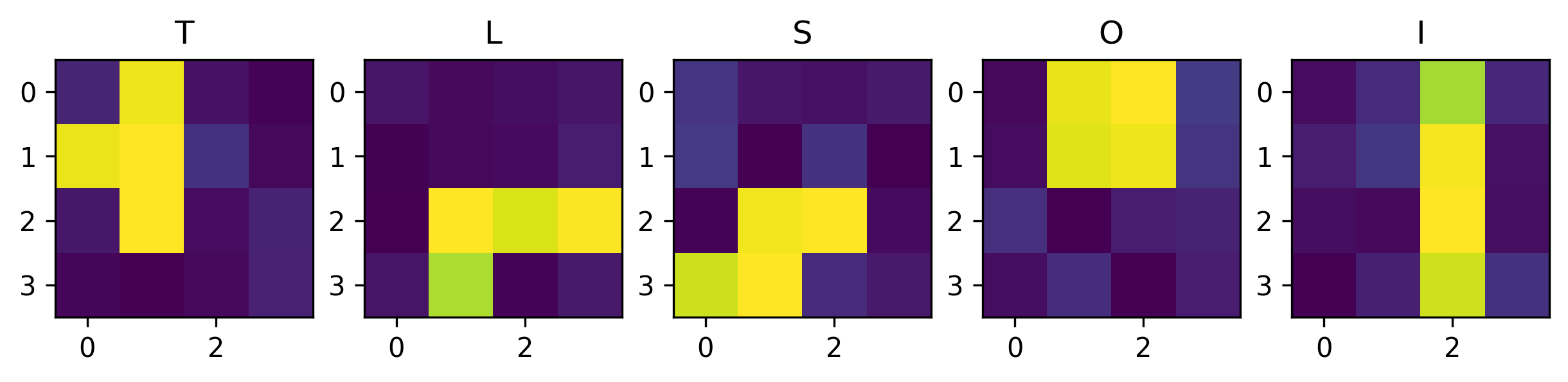}
    \caption{Visualization of instances of each class of the dataset \textit{Tetrominoes}.}
    \label{fig:tetrominoes}

\end{minipage}%
\hfill
\begin{minipage}[t]{0.3\textwidth}
    \vspace{0pt}
    \centering
    \includegraphics[width=\linewidth]{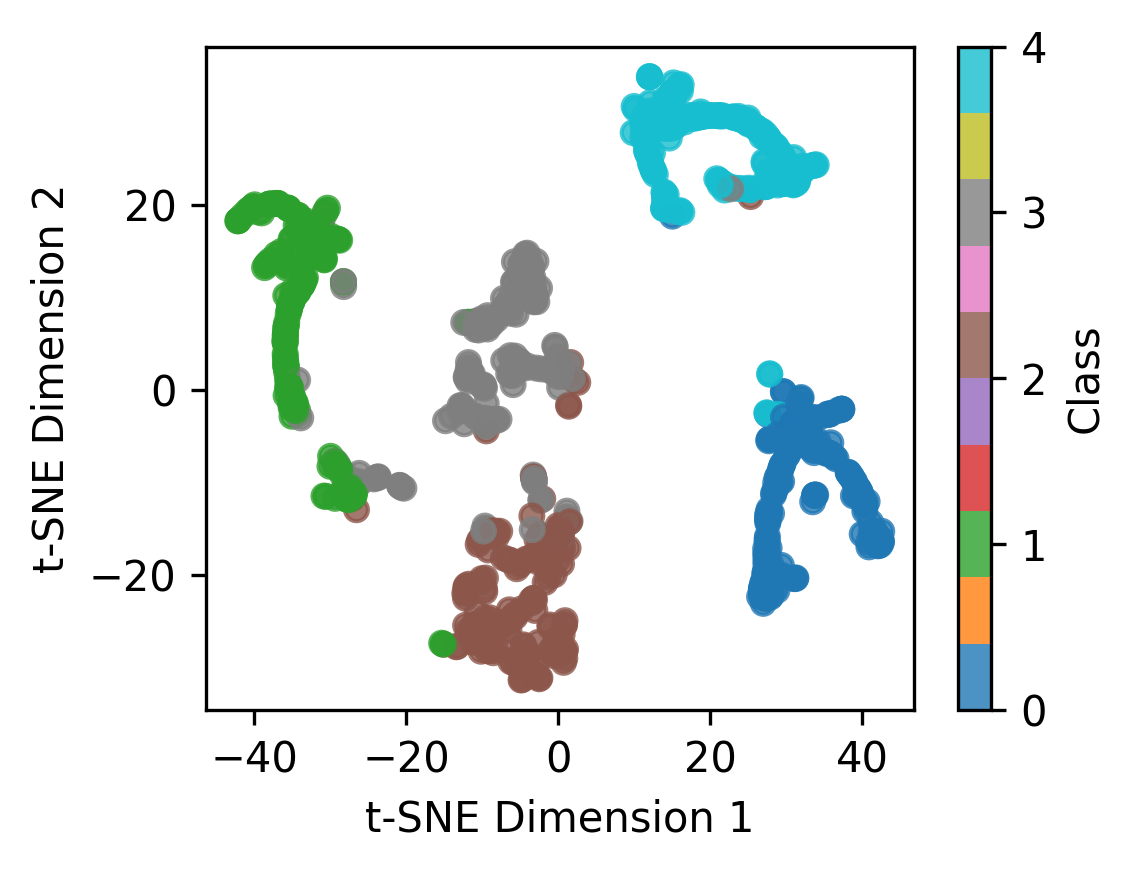}
    \caption{Visual representation of the dataset \textit{Blobs8}. The points were plotted along the first two t-SNE dimensions.}
    \label{fig:blobs5}

\end{minipage}
\end{figure*}

\subsection{Neural Collapse} \label{sec:NC}
Neural collapse is a recently discovered phenomenon that emerges in the terminal phase of training deep neural networks for classification tasks \cite{papyan_prevalence_2020}. First identified in the context of supervised learning with cross-entropy loss, Neural Collapse describes a set of remarkable and highly structured properties shared by the learned representations, classifier weights, and class means at convergence. This emergent geometric structure appears across a wide range of architectures, datasets, and optimization settings, suggesting that Neural Collapse reflects a fundamental principle underlying deep learning dynamics rather than an artifact of specific models or training choices \cite{papyan_prevalence_2020, lu_neural_2022,liu_leveraging_2024,ji_unconstrained_2022,zhu_geometric_2021,mixon_neural_2022}. Understanding Neural Collapse has therefore become an active area of research, offering insights into generalization, optimization, and the implicit biases of gradient-based learning.

This phenomenon is characterized by a vanishing variability within classes and the convergence of class centroids into a regular simplex Equiangular Tight Frame (ETF), among other properties. In other words, all data points belonging to the same class are embedded into the same class-mean vector in the latent space of the last layer, where each class-mean form a vertex of a simplex ETF. An ETF is a set of $k$ vectors in a $D$ dimensional space that are pairwise equiangular and maximally spaced \cite{fickus_equiangular_2017}. In fact, ETFs are structures that minimize the Welch bound

\begin{equation}
    \mu \leq \sqrt{\frac{k-D}{D(k-1)}},
\end{equation}

with $\mu = \max_{i,j} |\braket{v_i,v_j}|$, for any pair of vectors $v_i$ in the frame \cite{welch_lower_1974}. The regular simplex ETF corresponds to the special case where $k=D+1$, and therefore $\mu =\frac{1}{D}$. If the number vectors is equal to the dimension of the space $k=D$, then the ETF is formed by the orthogonal basis and $\mu=0$. For cases where $k>D+1$, the existence of ETFs is still an open area of research, having found results for specific cases, but without generalizable solutions for any combination of $k$ and $D$ \cite{sustik_existence_2007}. 

It has recently been proven by \cite{du_problem-dependent_2023} that QML models also exhibit the tendency to Neural Collapse. Authors show that when the empirical risk of quantum classifier approaches to zero ($\textrm{R}_{\textrm{ERM}} \rightarrow 0$), feature states and measurement operators tend to align with vanishing variability and that the class means tend to be mutually orthogonal. In fact, they prove that this structure is the global optimizer of the empirical risk. Since an orthogonal frame can be trivially turned into a regular simplex ETF by reducing its global mean, both classical and quantum Neural Collapse are equivalent.

These results are obtained for a regularized squared mean error loss, but in this work, we empirically analyze the tendency of quantum classifiers towards Neural Collapse of two additional losses. The results obtained by numerical experiments are presented in the following section.

\section{Results}

\subsection{Model description}\label{sec:model_description}

During this work, we trained models with different sets of observables, keeping the same structure in the parameterized quantum circuits. This way, any difference in the training or the performance of the model can be concluded to come from the choice of observables. On one hand, we studied projective measurements, described in sec. \ref{sec:projective_measurements}. From all possible options, sets of projectors into the computational basis states $\{\ket{i}\bra{i}\}_{i=0}^{K-1}$ were chosen for two main reasons. The first reason is that the computational basis is the default basis of measurement in quantum devices and the standard basis of the theoretical framework. The second reason is that this basis is an orthonormal basis, and since maximizing the expected values of the projector is equivalent to maximize the fidelity, training the QML with this set of observables allows us to enforce the orthogonality of the feature states, property happening in the Neural Collapse.

On the other hand, we studied sets of Pauli strings $\{P_i\}_{i=0}^{K-1}$, as described in sec. \ref{sec:paulistrings}. Unless explicitly stated otherwise, each set of Paulis was constructed from mutually commuting operators with the lowest possible locality. The motivation for this choice is twofold: mutually commuting Pauli operators can be measured simultaneously, thereby reducing the sampling complexity, and their common eigenbasis forms an orthogonal basis. 

In this setting, maximizing (or equivalently, minimizing) the expectation value of a Pauli operator maximizes the overlap between the evolved feature state and the positive eigenspace of that operator. In general, the positive eigenspaces of different Pauli observables have non-zero overlaps (see Fig. \ref{fig:eigenspaces}). However, it is possible to find a configuration of class mean state vectors where each centroid points into a region without overlaps, and therefore, achieve the ETF structure in the class means (see Fig. \ref{fig:eigencentroids}). This is only possible if the Pauli string commute and, consequently, QML models with mutually commuting Paulis are expected to reach Neural Collapse, whereas non-commuting sets of Paulis are expected to be unable to form an ETF structure in the class means.

Under this approach, a key distinction emerges between QML models trained with the two types of observables proposed in this work. When using projectors, an ETF structure can be enforced on a set of reference states, guiding the parameterized quantum circuit to map each data point to a region near its corresponding reference state. In contrast, when using Pauli observables, the ETF structure on class means is relaxed and remains only one of several possible solutions.

Two different loss functions where employed during this work. The first one is the \textit{cross entropy loss}, a widely used loss function in classification tasks. This loss measures the Kullback-Leibler Divergence between to probability distributions \cite{noel_alternate_2024}: the one-hot-encoded label distribution, and the predicted distribution. In our case, the predictions of the QML model formed a vector of expected values defined in \ref{eq:expval_vector}. In order to transform the entries of the prediction vector into normalized probabilities to belong a certain class ($p^k(x_i)$), a \textit{softmax} function was used:
\begin{equation}
    p^k(x_i) = \frac{e^{z_i^k}}{\sum_{i=1}^Ke^{z_i^k}} \quad \textrm{where} \quad z_i^k=\braket{O_k}_{\psi(\mathbf{x}_i,\theta)}.
\end{equation}

The second loss function, which we called \textit{fidelity loss}, was defined as:

\begin{equation}
    \ell_F(\mathbf{x}_i,y_i) = 1-|\braket{\psi(\mathbf{x}_i,\theta)|y_i}|^2+\lambda \sum_{\substack{j=1 \\ j \neq y_i}}^{K} |\braket{\psi(\mathbf{x}_i,\theta)|j}|^2.
\end{equation}

The factor $\lambda$ is a regularization term that penalizes the fidelity of the evolved feature state, with the non-label reference states. Note that when the number of classes $K$ is a power of two, due to the normalization of the quantum state vector, the loss takes the form of $  \ell_F(\mathbf{x}_i,y_i) = (1+\lambda)(1-|\braket{\psi(\mathbf{x}_i,\theta)|y_i}|^2)$. During the numerical experiments, there was no significant difference observed for different values of the penalization term, so $\lambda = 1$ was taken for the rest of the work.

As mentioned in \ref{sec:projective_measurements}, for pure states, the expected value of a projector equals the fidelity with respect to feature states, so one can write the fidelity loss in terms of expected values of projectors. Although replacing those fidelity terms with expected values of Paulis doesn't have the same geometric interpretation, models with Pauli observables were also trained using \textit{fidelity loss}, for sanity check. The summary of the main models trained during this work is presented in table \ref{tab:modeltypes}.

\begin{table}[h!]
\large
\centering
\begin{tabular}{c|c|*{2}{c|}}
  \multicolumn{2}{c}{} & \multicolumn{2}{c}{Loss Function} \\
  \cline{2-4}
  & & \bfseries Cross Entropy & \bfseries Fidelity \\[1ex]
  \cline{2-4}
  \multirow{2}{*}{\rotatebox[origin=b]{90}{Obs.}}
  & \bfseries Paulis & PauliCE & PauliF \\[1ex]
  \cline{2-4}
  & \bfseries Projectors & ProjCE & ProjF \\[1ex]
  \cline{2-4}
\end{tabular}
 \caption{Description of the main type of models trained during the work}
 \label{tab:modeltypes}
\end{table}

\subsection{Datasets and experiment setting} \label{sec:setting}

This section provides a brief overview of the datasets used in the numerical experiments and details the training of the parameterized quantum circuits. See table \ref{tab:datasets} for the description of the datasets.
\begin{table}[h]
\centering
\small 
\begin{tabular}{|c|p{2cm}|c|c|c|}
\hline
\textbf{Name} & \textbf{Description} & \textbf{Features} & \textbf{Classes} & \textbf{Size} \\ \hline
\textit{Blobs2} &
Random points in the plane, drawn from a Gaussian distribution around the centroids (see Fig. \ref{fig:blobs3}). &
2 & 3 & 10000 \\ \hline
\textit{Blobs8} & Random points in a 8-dimensional space, drawn from a Gaussian distribution around the centroids (see Fig. \ref{fig:blobs5}) & 8 & 5 & 10000 \\ \hline
\textit{Tetrominoes} & 4x4 images of tetrominoes in grayscale (see Fig. \ref{fig:tetrominoes}) & 16 & 5 & 400 \\ \hline
\textit{Panels} & Image patches of solar panels with cracks (see Fig. \ref{fig:panels})& 5476 & 3 & 368\\ \hline
\end{tabular}
\caption{Description of used datasets.}
\label{tab:datasets}
\end{table}

All quantum circuits were programmed and run in ideal (noiseless) state vector simulators using \textit{Pennylane} library \cite{bergholm_pennylane_2022}, along with \textit{JAX}'s interface \cite{jax2018github} to speed up the simulations. Angle encoding method \cite{schuld_machine_2021} was followed to embed classical data into the Hilbert space, employing one qubit per feature. For the variational part, \textit{StronglyEntanglingLayers} template was used with a single layer repetition. The sequence of encoding layer and variational layer was repeated $n_l$ times, following the \textit{data re-uploading} method, in order to make the model generate nonlinearities of the features \cite{perez-salinas_data_2019}. Hence, the total number of parameters in the models is $n_{\textrm{params}} = 3n_qn_l$. The update of the parameters was made using an \textit{Adam} optimizer \cite{kingma_adam_2017} with a cherry-picked learning rate for each dataset, but equal between different models.

For the \textit{Panels} dataset, the images were too high-dimensional to embed directly in a simulated quantum circuit. A classical convolutional neural network (CNN) was therefore employed to reduce the feature dimensionality before inputting them into the quantum circuit. The classical parameters were trained jointly with the variational quantum parameters for each model, using the same CNN architecture across all models. For the evaluation of the classification performance, the \textit{F1} score was used with \textit{macro averaging}, a score considered to be sensitive to class imbalances \cite{tharwat_classification_2020}. All datasets where split into a training and testing set, randomly, using a ratio of $3/4$.

\begin{figure*}[t]

    \begin{subfigure}{0.45\textwidth}
        \centering
        \includegraphics[width=\linewidth]{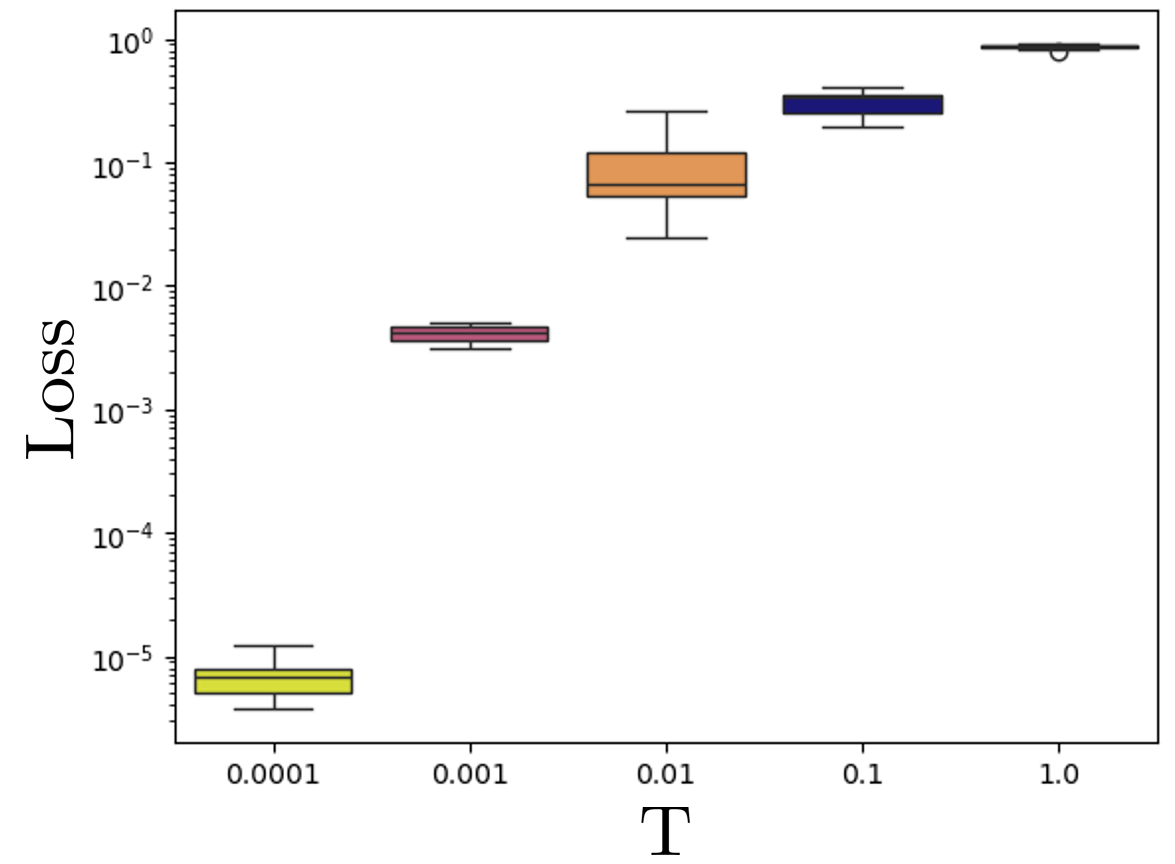} 
        \caption{Loss reached in the last epoch, in function of the temperature factor in \textit{softmax}.}
        \label{fig:temploss}
    \end{subfigure}
    \begin{subfigure}{0.45\textwidth}
        \centering
        \includegraphics[width=\linewidth]{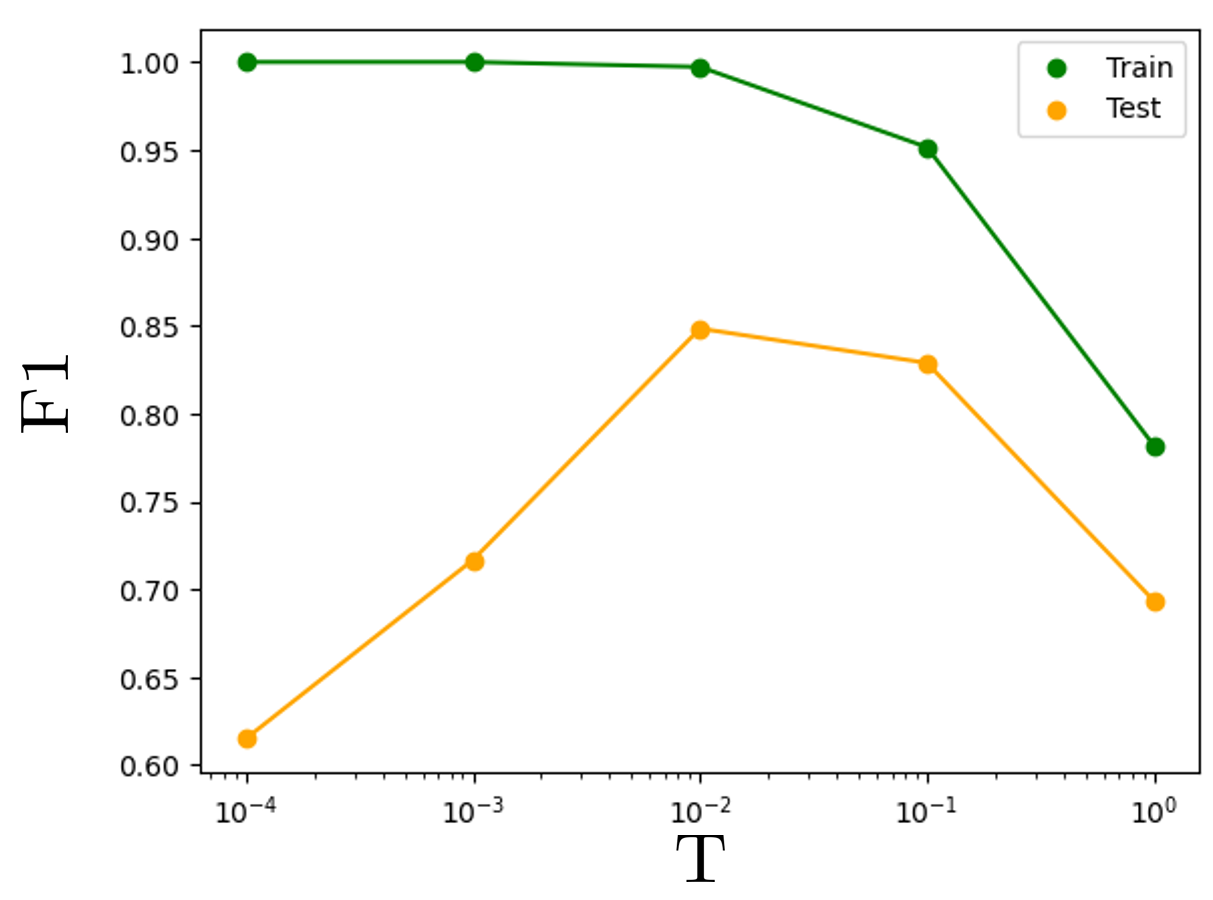}
        \caption{Mean $F1$ score in function of the temperature factor in \textit{softmax}.}
        \label{fig:tempf1}
    \end{subfigure}

    \caption{The effect of the temperature factor in the loss and $F1$ score of a \textit{PauliCE} model. For each value of $T$, 5 randomly initialized models were trained, with $n_l=5$ and $100$ epochs. Models trained for \textit{Tetrominoes} dataset.}
    \label{fig:temperature}
\end{figure*}

\subsection{Numerical results}\label{sec:numerical_results}

First, and before starting to compare different model types, a subtle detail was noted. The \textit{softmax} function, used to transform model outputs $z_i^k$ into a normalized probability distribution, takes input values from a domain $(-\infty,\infty)$. However, the expected values of Pauli operators are bounded in $[-1,1]$, which makes all \textit{softmax} values fall close to its central value $0.5$ and the models be unable to reach 0 \textit{cross entropy loss}. For this reason, a \textit{temperature} term was included to rescale the outputs of the model $z_i^k \rightarrow \frac{z_i^k}{T}$.

Fig. \ref{fig:temperature} shows how the temperature term affected the minimum loss the models could achieve in the training and the $F1$ classifications score. Decreasing the temperature made $z_i^k$ values rescale to a wider domain of the \textit{softmax} function, and consequently, the models approach better the 0 of the cross entropy loss. On the other hand, this simple rescaling revealed a significant change in the classification power of the models: Decreasing the temperature improved the classification in the training set until a perfect classification, but the classification in the test set showed an initial improvement and a degradation after a maximum value. This behavior suggests a tendency to \textit{overfitting} \cite{lopez_overfitting_2022}. Accordingly, the temperature of the \textit{softmax} can be used as a regularization hyperparameter that can affect to the generalization ability of the model. For the rest of the work $T=0.01$ was chosen, since it was the value maximizing the classification score in the test set, but as a good practice, a rigorous parameter search should be made for each dataset and trained model.

For the study of Barren Plateaus, the concentration of the loss was studied by randomly sampling points of the multidimensional parameter space. For different values of re-uploading layers $n_l$ and locality $k$ of the observables, 100 variational parameter vectors where randomly sampled where every vector element $\theta_i\in[0, 2\pi)$ was drawn from a uniform distribution, and the loss function was evaluated for each parametrization of the quantum circuit\footnote{Generating uniformly random quantum unitaries requires sampling according to the Haar measure. Determining the corresponding parameter distributions for a given parameterized circuit, however, is generally difficult and becomes intractable in high-dimensional settings. As a result, most quantum machine learning applications rely on sampling circuit parameters from simple distributions, such as uniform distributions, which do not in general reproduce the Haar measure.}.

\begin{figure*}[h]
    \begin{subfigure}{0.45\textwidth}
        \centering

        \includegraphics[width=\linewidth]{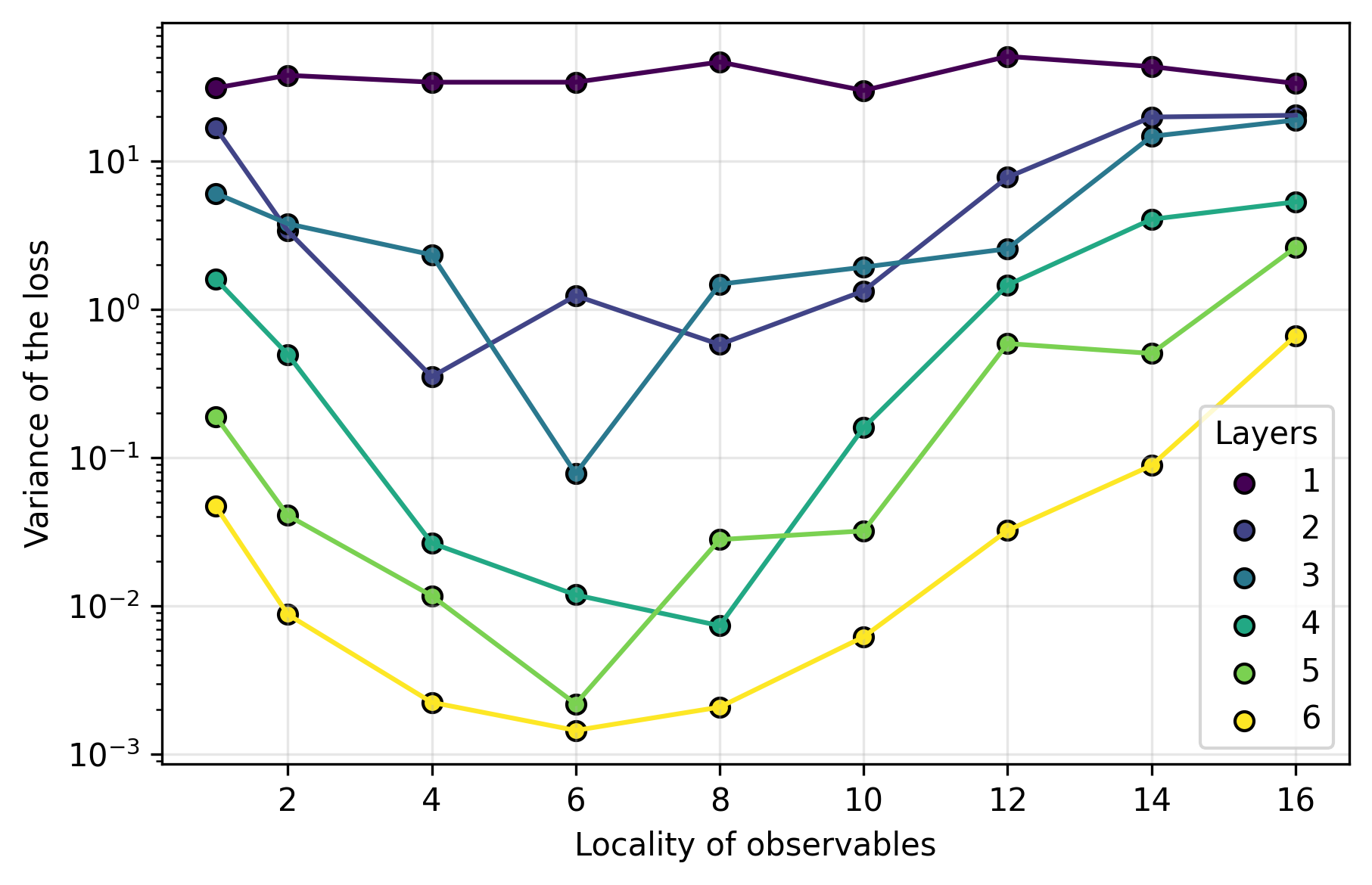} 
        \caption{Paulis}
        \label{fig:loss_landscape_pauli}
    \end{subfigure}
    \begin{subfigure}{0.45\textwidth}
        \centering

        \includegraphics[width=\linewidth]{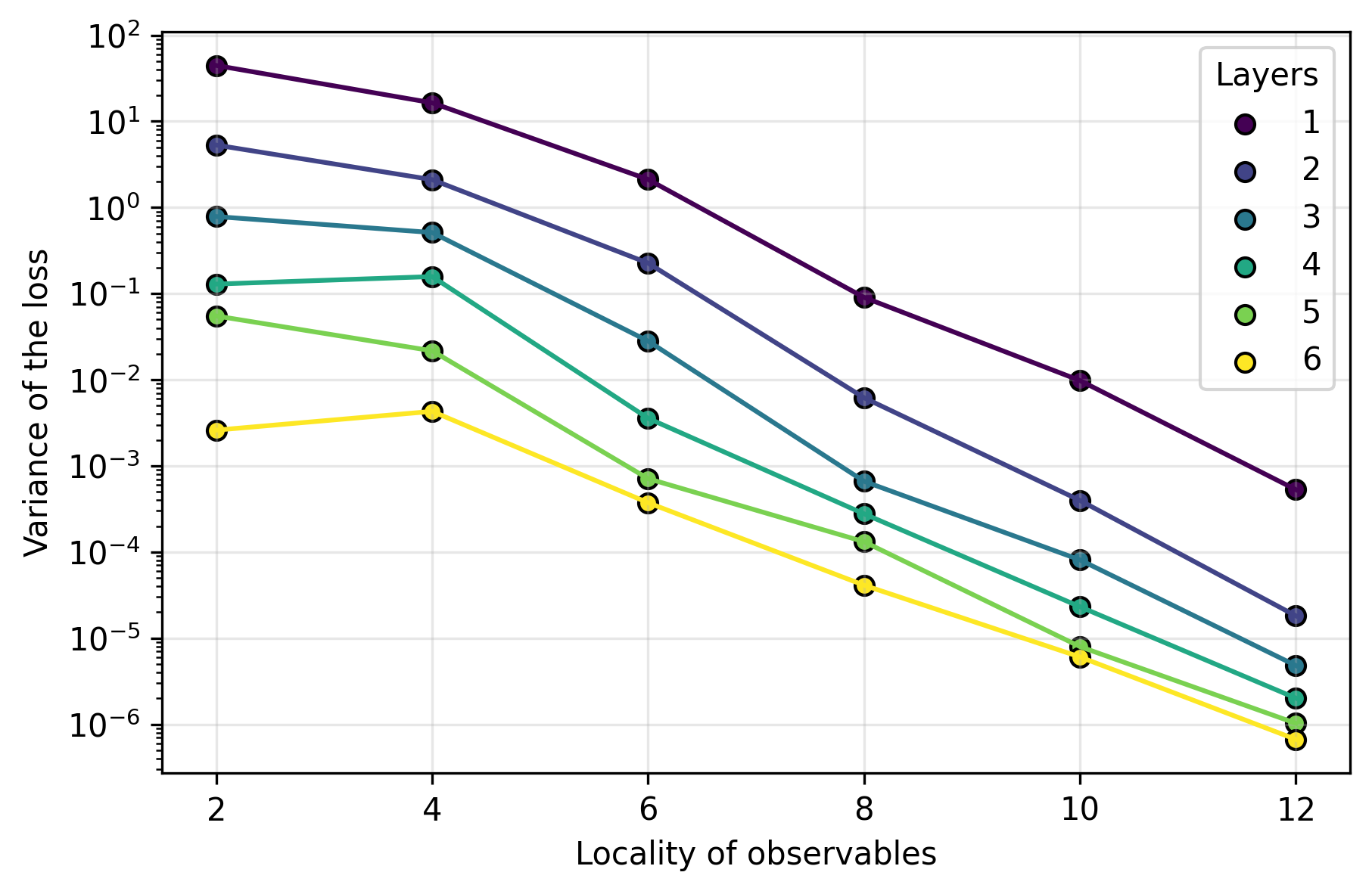}
        \caption{Projectors}
        \label{fig:loss_landscape_proj}
    \end{subfigure}
    
    \caption{Variance of the loss in function of the locality of observables, across randomly initialized variational parameters. Points of different colors represent parametereized quantum circutis with different data re-uploading layers $n_l\ =1,\dots,6$.  Dataset: \textit{Tetrominoes} (3 classes). }
    \label{fig:temperature}
\end{figure*}

In the case of Pauli observables, a set of $\{X^{\otimes k},Y^{\otimes k},Z^{\otimes k}\}$ operators was chosen to classify 3 classes, where $k$ is the locality of the Pauli string. Fig. \ref{fig:loss_landscape_pauli} shows a clear dependency of the variance of the loss function w.r.t. the locality of the Pauli strings. However, this dependency was not monotonically decreasing, as one could initially expect from previous results following the operator causal cone approach \cite{uvarov_barren_2021}. On the contrary, the dependency of the loss showed a parabolic shape, even showing higher variance with high-locality observables than with low-locality observables. 

This behavior can be explained recalling the Lie-algebraic expression of the variance of the loss, introduced in \ref{sec:BPs}. If we divide the space linear operators $\mathcal{B}$ acting on $n_q$ qubits into a direct sum of modules with different localities $\mathcal{B}=\bigoplus_k^{n_q}\mathcal{B}_k$, we can express the variance in terms of a sum of contributions of each module \cite{diaz_showcasing_2023}:
\begin{equation}
\label{BPmodules}
 \operatorname{Var}_{\theta}[l_\theta(\rho,O)] = \sum_{k=1}^{n_q} \frac{\mathcal{P}_{\mathcal{B}_k}(\rho)\mathcal{P}_{\mathcal{B}_k}(O)}{\textrm{dim}({\mathcal{B}_k})}.
\end{equation}
Since the observables have a well defined locality, the purity term $\mathcal{P}_{\mathcal{B}_k}(O)$ is nonzero only when the index $i$ matches the locality $k$. In consequence, only the module with the same locality as the observable is making a contribution to the variance of the loss. If we study the dimension of the module of locality $k$ we see that it grows with $\binom{n_q}{k}$, a function that is symmetrical with respect to $k\rightarrow  n_q-k$ and a minimum around $n_q/2$. This explains the parabolical shape of the variance of the loss and the asymmetries come from the purity term of the evolved state.

Additionally, increasing the number of re-uploading layers in the parameterized quantum circuit also decreases the variance of the loss. The reason behind this is that the Lie-algebraic expression \ref{BPmaster} assumes the parameterized quantum circuit to form at least a 2-design of the unitary group $G=e^{\mathfrak{g}}$, in order to approximate its statistics up to the second moment, but shallow quantum circuits do not satisfy this condition \cite{ragone_lie_2024}. Adding more layers generates more elements of the theoretical DLA and makes the quantum circuit approximate better the statistics of the unitary group.

When it comes to projectors, the locality of such operators is not that well defined as it is in the case of Pauli strings. For instance, as Pauli strings form a basis of the space of Hermitian operators, we can write the projectors of computational basis states in terms of Pauli strings:

\begin{equation}
    \ket{i}\bra{i}=\bigotimes_{j=1}^{n}\frac{1}{2}\big(I+(-1)^{i_j}Z_j \big)
\end{equation}

where $i_j$ are the bits of the bit-string $i=i_{n}i_{n-1}\dots i_2i_1$ and $Z_j$ is the local Pauli matrix acting on qubit $j$. For this expression, we can see that projectors can be written as a linear combination of Pauli strings with mixed localities. For the experiment represented in Fig. \ref{fig:loss_landscape_proj}, the locality of projectors was defined as \textit{maximum locality of the Pauli string in its linear combination}, or equivalently, the number of measured qubits $n$, or the length of the bit-string. Fig. \ref{fig:loss_landscape_proj} show how, unlike Pauli observables, projectors show a monotonic exponential concentration of the loss, since these projectors live in a modulus of operators of mixed localities, that grow exponentially in size with the number of qubits they act on. Again, depending on the increase of re-uploading layers $n_l$, the same behavior as with Pauli strings was observed in the variance of the loss.

For the study of Neural Collapse, two indicators where computed:
\begin{itemize}
    \item \textit{Intra}: Mean fidelity between the evolved feature states with their class means.
    \item \textit{Inter}: Mean fidelity between class mean state vectors (centroids).
\end{itemize}

Models trained until the terminal phase are expected, upon reaching Neural Collapse, to maximize the \textit{Intra} indicator—causing all data points to converge to their respective class means in the feature space—while minimizing the \textit{Inter} indicator, resulting in class mean vectors arranged in a (simplex) ETF structure.

\begin{figure*}[h]
    \centering
    \includegraphics[width=0.9\linewidth]{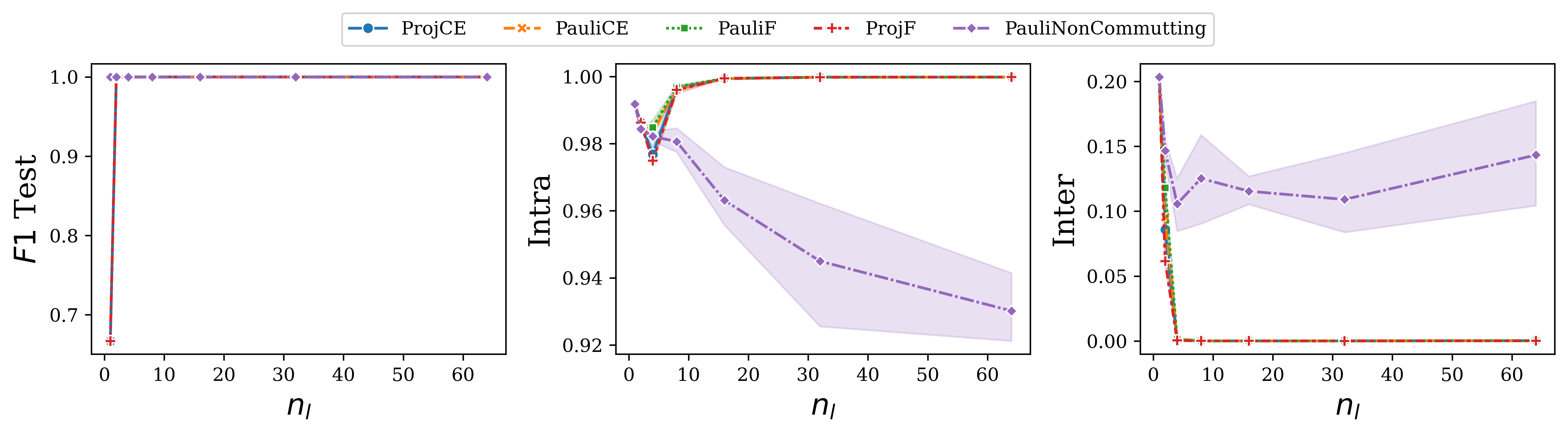}
    \caption{Comparison of Neural Collapse indicators of different model types. F1 classification score, \textit{Intra} and \textit{Inter} indicators in function of the re-uploading layers $n_l$. Models were trained for the dataset \textit{Blobs2}, running 100 epochs in the training, and 5 random initializations for every model and value of $n_l$.}
    \label{fig:exp1}
\end{figure*}

Fig. \ref{fig:exp1} show these Neural Collapse indicators for each of the model types introduced in \ref{sec:model_description}. Additionally, a new model type called \textit{PauliNonCommuting} was included with a set of observables made of non-comutting Paulis, e.g. $\{X,Y,Z\}$. Results show that, up from a small number of re-uploading layers, all models are able to achieve perfect classification, which is an expected result considering the simplicity of the problem. Nonetheless, increasing the re-uploading layers $n_l$ made the models approximate better the Neural Collapse by maximizing the within class fidelity and minimizing the fidelity between class centroids. In the case of \textit{PauliNonComutting} models, however, models were not able to reach the neural collapse due to the non-zero overlap of eigenspaces of non commuting operators. 
Studying the evolution of \textit{F1}, \textit{Intra} and \textit{Inter} indicators through the training process, it was observed that Neural Collapse indicators kept optimizing even after the models surpassed the point of perfect classification, coinciding with the fact that Neural Collapse happens in the terminal phase of training.

\begin{figure*}[h]
    \centering
    \includegraphics[width=0.9\linewidth]{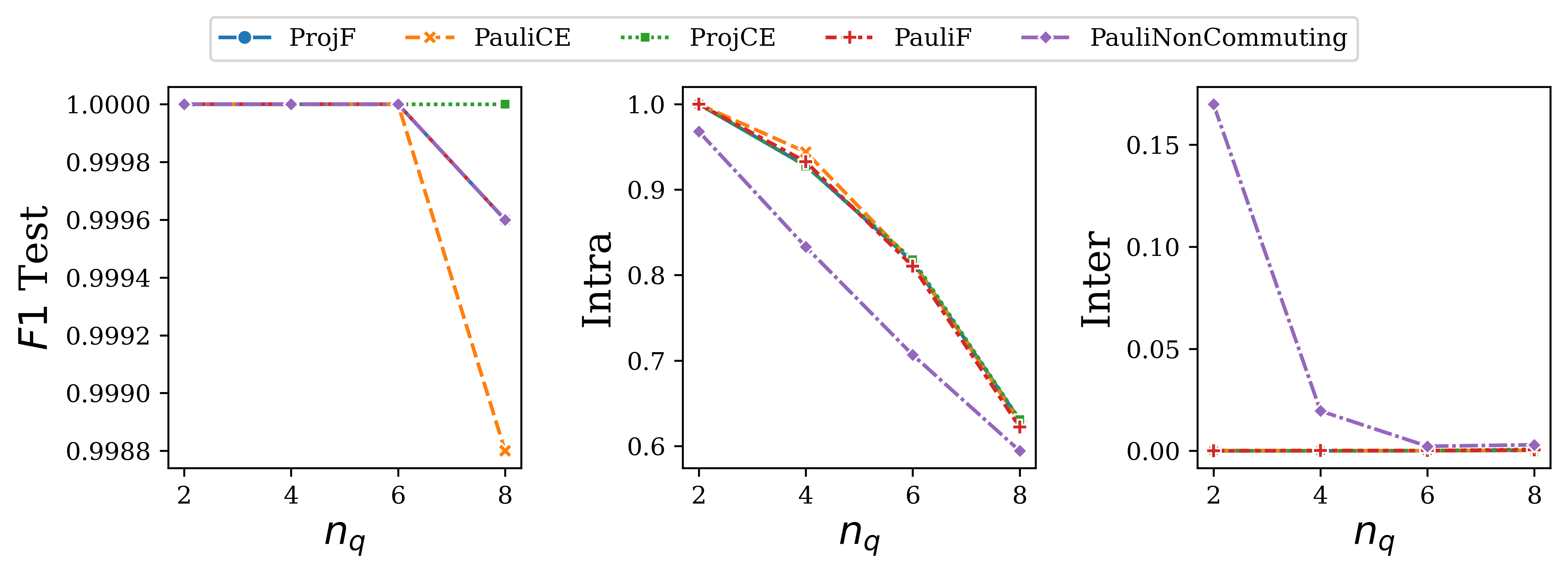}
    \caption{Decrease in Neural Collapse indicators with the number of qubits $n_q$ in the parameterized quantum circuit. Consequence of the course of dimensionality. Models were trained for the dataset \textit{Blobs2}, running 100 epochs in the training. }
    \label{fig:curse_of_dimensionality}
\end{figure*}

In the next experiment, the same model types were trained for the same \textit{Blobs2} dataset, but this time, the number of qubits in the parameterized quantum circuit was increased gradually. The results showed that fidelities (both \textit{Intra} and \textit{Inter}) decreased with the number of qubits (see Fig. \ref{fig:curse_of_dimensionality}). This behavior can be explained as a consequence of the curse of dimensionality \cite{verleysen_curse_2005}. As the dimension of the Hilbert space increased, the feature states became more sparse, decreasing the fidelity between them. Indeed, the probability distribution of the Fidelity $F$ between two Haar-random (pure) quantum states is given by

\begin{equation}
    P(F)=(d-1)(1-F)^{d-2}
\end{equation}

with $d=2^{n_q}$ the dimension of the Hilbert space. From there, we can compute that the expected value of the fidelity $\mathbb{E}[F]=\frac{1}{d}$ decreases exponentially with the number of qubits. This fact brings two consequences when it comes to Neural Collapse. First, this makes the convergence to class centroids more difficult for the models, requiring more training epochs and/or model complexity to reach the Neural Collapse. Apart from that, the curse of dimensionality makes feature states, and in consequence class centroids, to be mutually orthogonal, vanishing differences in the capacity of models of  achieving ETF structure in their class centroids.

Previous experiments showed how quantum machine models can reach the Neural Collapse using both sets of projectors and (commuting) Pauli operators. However, models that have enough capacity (model complexity) to reach the Neural Collapse are expected to be in the over-parameterization regime \cite{dar_farewell_2021, larocca_theory_2021}, unless they have a strong inducted bias. In realistic scenarios, quantum machine learning models will rarely be in that over-parameterization regime, and the models wont usually be trained until the terminal phase of training. Taking this into account, a sensible question can arise: \textit{Does imposing an ETF structure in the set of observables bring any benefits in the training or the performance of under-parameterized quantum machine learning models?}

 \begin{figure*}[h]
    \centering
    \includegraphics[width=0.9\textwidth]{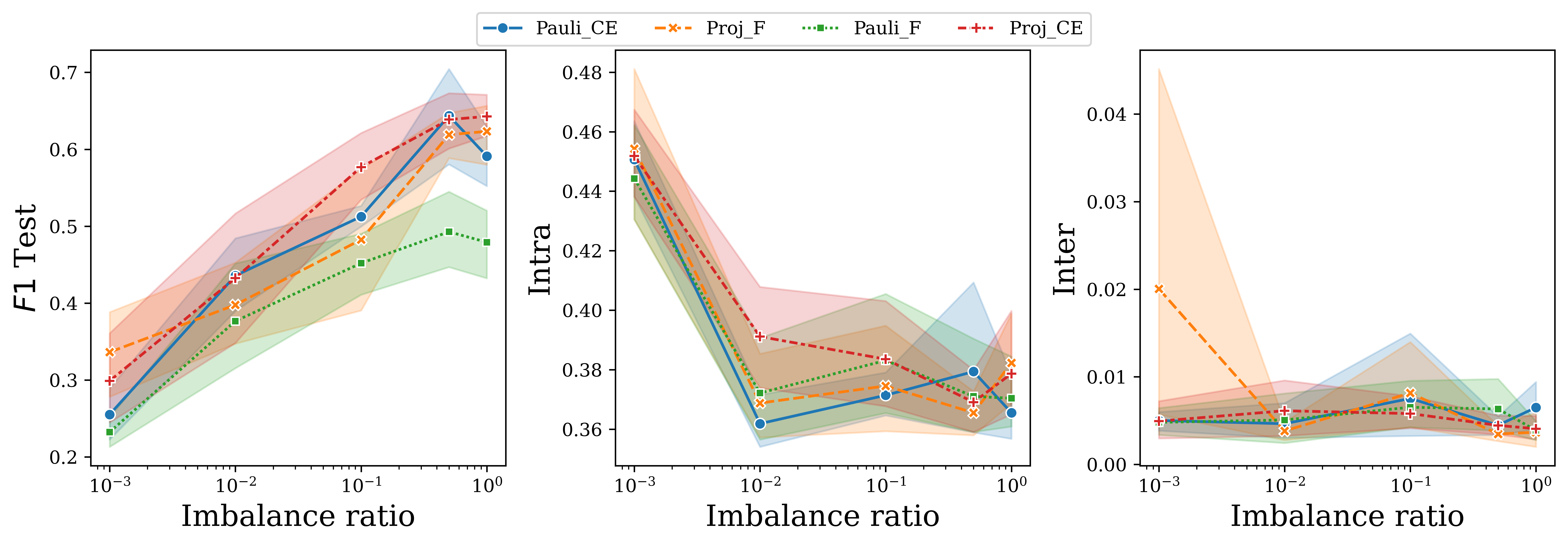}
    \centering
    \caption{F1 score and Neural Collapse indicators in function of the imbalance ratio in the dataset \textit{Blobs8}. 10 models of each type were trained, for 100 epochs and $n_l=5$ re-uploading layers.}
    \label{fig:imbalance}
\end{figure*}

 There are seminal works that demonstrate that fixing the last-layer weights of neural networks to an ETF reduces memory usage and improves the performance on imbalanced datasets and few shot learning tasks \cite{xu_quantifying_2023, yang_inducing_2022, thrampoulidis_imbalance_2022, yang_neural_2023}. If the answer of the previous question is affirmative, we could expect to observe a difference between models with Paulis and Projector, since Projectors do impose ETF structure, while Paulis are more flexible in that aspect (see sec. \ref{sec:model_description}). Two more experiments were driven with the goal of capturing those differences between sets of observables.

\begin{figure}[h]
     \centering
     \includegraphics[width=\linewidth]{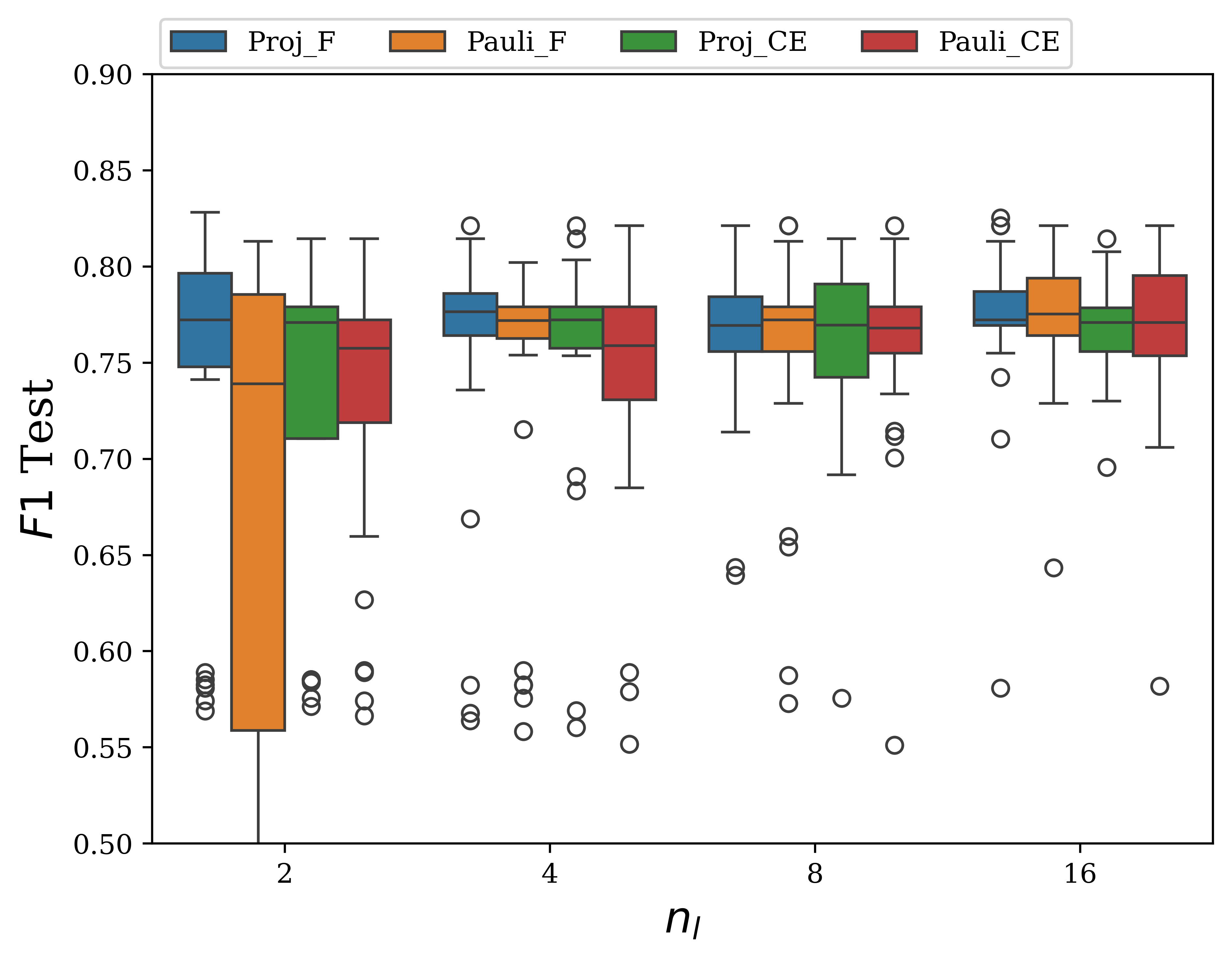}
     \caption{Classification results of the main model types with different re-uploading layers $n_l$, trained for \textit{Panels} dataset (see table \ref{tab:datasets}). 30 models of each type were trained for each value of $n_l$, all of them trained in 100 epochs.}
     \label{fig:exp9}
 \end{figure}

 In the first experiment, several models were trained in a real-world classification setting: defect classification with solar panel images. As explained in sec. \ref{sec:setting}, classical convolutional layers were employed to reduce the dimensionality of the images, to be encoded afterwards in the parameterized quantum circuits. Results in \ref{fig:exp9} show that models with projectors as observables performed slightly better, on average, for circuits with low number of re-uploading layers. Adding re-uploading layers, however, made those differences vanish and all models show similar performances. 

For the second experiment, the synthetic dataset \textit{Blobs8} (see table \ref{tab:datasets}) was created. Afterwards, to create an imbalanced scenario, all classes except one were randomly under-sampled, keeping the total size of the dataset constant with 10000 samples. This way, subsets with different ratios of imbalance (size of minority class / size of majority class) were created. Fig. \ref{fig:imbalance} show the classification performances of each model type, in function of the imbalance ration in the dataset. All models showed a gradual degradation of the performance as the imbalance ratio decreased. Models with projectors showed slightly better results than models with Pauli strings, but the differences were not significant enough to conclude that there exists a clear advantage coming from the imposition of ETF structure in the observables.

\section{Conclusion} \label{sec:conclusion}

In this work, we have studied the problem of multiclass classification using parameterized quantum circuits as quantum machine learning models. More precisely, two main approaches have been studied: maximizing the expected value of a set of quantum observables, on one hand, and fixing a set of reference quantum states to make the data-encoded quantum states maximize the fidelity with their respective class vectors, on the other hand. As an example of the first approach, sets of Pauli strings have been chosen; while for the second approach, sets of projectors into the computational basis states have been selected.

Models with both types of observables have been trained using two different loss functions: \textit{cross entropy} and \textit{fidelity} loss functions. The training of the models has been performed in noiseless simulators using four different datasets, including one with real-life solar panel images.

During the experimentation, first, the problem of Barren Plateaus has been studied empirically by observing the concentration of the loss as a function of the locality of the observables in the sets. The results have shown that while projectors present a monotonic exponential descent of the variance of the loss, Pauli strings have exhibited a parabolic behavior, recovering large variances for highly local strings.

Afterwards, the phenomenon of Neural Collapse has been studied for each type of model. We have postulated that sets of projectors can be used to impose the ETF structure in the class-mean quantum state vectors, whereas Pauli strings are more flexible in the display of the class-mean vectors. In order to observe this phenomenon experimentally, two indicators have been defined: intra-cluster fidelity and inter-cluster fidelity indicators. Experimental results have shown that (at least in low-dimensional Hilbert spaces) both models with projectors and Pauli strings have been able to reach Neural Collapse by maximizing the intra-class fidelity and minimizing the inter-class fidelity. On the contrary, results have demonstrated how sets of non-commuting Pauli strings make a model unable to form the necessary ETF structure in the class-mean vectors due to the overlap of eigenspaces of the observables. In high-dimensional Hilbert spaces, however, as a consequence of the curse of dimensionality, the class-mean vectors have become mutually orthogonal, vanishing the difference in the capacity of reaching Neural Collapse between commuting and non-commuting sets of Pauli strings.

Since Neural Collapse is expected to happen in over-parameterized models in the terminal phase of training, two experiments have been conducted to try to observe whether imposing ETF structure in class-mean vectors could bring benefits in the performance of under-parameterized quantum machine learning models. For these, models have been trained on a real-life dataset and a synthetic imbalanced dataset with different ratios of imbalance. Results have shown that, for a low number of re-uploading layers in the quantum circuit, projectors have performed slightly better than Pauli strings, but the differences have not been significant enough to confirm or deny the hypothesis. Apart from these main results, we have experimentally observed how the temperature scaling factor in the \textit{softmax} function can act as a regularizer, controlling the generalization capacity of a model.

Finally, we have identified as an interesting direction for future research the theoretical and empirical investigation of how enforcing specific geometric structures on the class-mean vectors in Hilbert space may provide advantages in quantum machine learning, analogous to those observed in classical deep learning.

The theoretical overview, experimental results, and conclusions presented in this work are expected to help other researchers gain a deeper understanding of the dynamics of multiclass learning with parameterized quantum circuits, contribute insights into the engineering of quantum machine learning models, and serve as a foundation for further studies in this area.

\section{Acknowledgments}
This article has been supported by the Government of the Basque Country through the research grant ELKARTEK KUBIBIT - Grant number KK-2025/00079 (Kuantikaren Berrikuntzarako Iibilbide Teknologiokak). Dataset \textit{Panels} has been provided by \textit{Mondragon Assembly}.

\printbibliography

\EOD
\end{document}